\documentclass[11pt,a4paper]{article}

\usepackage[utf8]{inputenc}
\usepackage[T1]{fontenc}
\usepackage{lmodern}
\usepackage{amsmath}
\usepackage{amssymb}
\usepackage{amsthm}
\usepackage{booktabs}
\usepackage{longtable}
\usepackage{array}
\usepackage{microtype}
\usepackage[margin=1in]{geometry}
\usepackage[colorlinks=true,linkcolor=blue,citecolor=blue,urlcolor=blue]{hyperref}
\theoremstyle{definition}
\newtheorem{definition}{Definition}

\title{Compiling Sufficient Governance Context from Declared Losses and Reachable States\\[0.3em]\large Exact Observation-Contract Synthesis with Cardinality and Cost Objectives}
\author{Gaston Besanson\thanks{Universidad Torcuato Di Tella}}
\date{}

\begin{document}
\maketitle

\noindent Companion artifact: \texttt{sarc-authority-derivation}. Paper 5 of the SARC series, built on the pinned \texttt{sarc-suite-one-pass} artifact (arXiv 2608.18360, commit \texttt{782261e}) as a read-only imported baseline.

\begin{abstract}
We call the object this paper derives and certifies a \textbf{minimal sufficient governance context}: given a finite reachable-state model, a deterministic declared verdict, and a set of candidate observable attributes, we compute sufficient observation sets, distinguish attributes that are individually indispensable from contracts that are jointly sufficient, and select among sufficient contracts under a cardinality or a declared observation-cost objective. An \textbf{observation contract} is a set of candidate attributes whose values determine the declared verdict on every reachable state; an \textbf{authority contract} is an observation contract selected under one of those objectives and bound to a gate schema. We do not declare an observation contract and hope it is complete: we \emph{synthesize} every inclusion-minimal sufficient contract where exhaustive enumeration is affordable, and a minimum-cardinality or minimum-cost sufficient contract by SAT/MaxSAT encoding otherwise; for every selected contract reported in the evaluation, we separately check sufficiency. On a preregistered constructed code/cloud domain (not planted for this purpose), the individually-indispensable core is \textbf{not sufficient} as an observation contract (six of ten candidates) and two distinct seven-attribute reducts exist; a preregistered declared-cost model separates them exactly (CH-B2: \textbf{supported}, the cheaper contract costs 6.124 against 7.354 for the alternative, both independently checked sufficient before cost is consulted). On a second, independently structured, thirty-five-property constructed domain, preregistered before any core, reduct, or cost outcome was computed, the same core-insufficiency pattern recurs (CH-C1: \textbf{supported}, a four-property core, minimum contract cardinality nine, at least five distinct minimum-cardinality contracts) -- but the same domain's own cost model does not separate them: CH-C2 is \textbf{not supported}, a fully explained cost tie across every known minimum-cardinality contract at 9.153, reported as found. We measure discernibility-family scaling on candidate-attribute universes up to one hundred properties, where exhaustive enumeration is expected and confirmed infeasible within a registered three-hundred-second timeout under this project's own measured setup, while SAT/MaxSAT synthesis solves in well under a second on the same instances; MaxSAT showed no measured cardinality advantage over plain SAT on these specific families. AuthorityBench compares four baselines -- a declared-only manual policy, ABAC-mining reimplemented from Xu and Stoller (2015), essential-variable analysis, and exhaustive reduct enumeration -- across three domains; the declared-only baseline is not exactly sufficient on any of them. Every selected contract is checked for sufficiency, with a concrete counterexample returned on failure and a check summary (a partition-cell count and a uniformity flag, not a portable certificate) on success; this is the compiler-focused scope of the two-scope table in Section~\ref{sec:conclusion} -- a separately coded, independently specified end-to-end case study is registered follow-up work, not claimed here.
\end{abstract}

\section{Introduction}
\label{sec:intro}

\noindent\textbf{The decision problem.} A pre-action authority gate is exactly as complete as whoever declared its observation set thought to be. This artifact's own imported baseline (\texttt{specs/authority.yaml}) observes two things: whether the acting role is on an allow-list, and whether a declared order value exceeds a single scenario-wide cap. Acquiring every candidate field instead of choosing is not free either: latency, privacy exposure, and availability each carry a cost a gate pays on every decision. Between ``declare an observation set by hand and hope it is complete'' and ``observe everything,'' this paper compiles the third option: derive, from a declared loss model and a declared reachable-state model, the exact observation contracts sufficient to reproduce that model's verdict, and select among them by an explicit objective.

\noindent\textbf{Running example.} Section~\ref{sec:results}'s own CH-B1 result gives the concrete shape of the problem before any formalism: on a preregistered constructed code/cloud execution-agent domain, the core (Section~\ref{sec:formal}) is \textbf{not sufficient} as an observation contract on its own, and there are exactly two operationally different ways to close the remaining gap -- the core plus \texttt{branch}, or the core plus \texttt{environment}, each independently checked sufficient. Neither is more ``correct'' than the other on cardinality alone; Section~\ref{sec:results} shows a declared cost model can break the tie.

\noindent\textbf{Contribution statement.} This paper separates three things, and claims no more for any of them than its own evidence supports: first, a formulation of governance-context compilation as sufficient-set synthesis over a declared loss model and a declared reachable-state model (Section~\ref{sec:formal}); second, an implemented pipeline -- discernibility construction, SAT/MaxSAT synthesis, sufficiency checking -- exposed through one packaged entry point (Sections~\ref{sec:method}--\ref{sec:implementation}); and third, a bounded evaluation across constructed domains, measuring semantic correctness, contract multiplicity, cost separation, and solver scaling (Sections~\ref{sec:evaluation}--\ref{sec:computational}). Test counts, correction history, and engineering-gate counts are not scientific contributions and are not presented as such.

\noindent\textbf{Guarantee boundary, stated once here and held to throughout.} Correctness in this paper is relative to a declared loss model, a declared candidate-attribute representation, and a declared reachable-state set -- never a general safety guarantee independent of those three declarations.

\noindent\textbf{Scope.} This is the compiler-focused paper of the two legitimate scopes Section~\ref{sec:conclusion} distinguishes: formulation, implementation, and a bounded evaluation against declared, constructed models, not an end-to-end operational validation. A case study with domain-owner-specified losses, an equal-information baseline comparison, and a separately coded reference executor is registered future work (Section~\ref{sec:supplementary}), not performed and not claimed here.

\section{Problem Formulation and Guarantee Boundary}
\label{sec:formal}

\noindent\textbf{Inputs.} A finite reachable set $R$ of executable-reachable state tuples over a declared set of candidate attributes $A$; a deterministic declared verdict $v_M$, a boolean function of one reachable tuple; and, optionally, a strictly positive additive declared observation cost $c(a)$ per candidate attribute.

\noindent\textbf{Representation assumption.} The full candidate-attribute projection of a reachable tuple must determine $v_M$'s verdict on it -- a consistency requirement on the declared model, not an assertion that observing every candidate attribute suffices for an arbitrary malformed or incomplete input.

\begin{definition}[Observation contract]
An observation contract is a set $S$ of candidate attributes. $S$ is \textbf{sufficient} for $v_M$ on $R$ (Definition~4) iff any two reachable tuples agreeing on every attribute in $S$ also agree on their $v_M$ verdict. A sufficient observation contract selected under an explicit objective (minimum cardinality or minimum declared cost) and bound to a gate schema is an \textbf{authority contract}.
\end{definition}

\begin{definition}[Participation, the one-flip criterion]
Attribute $a$ is authority-bearing for $v_M$ iff there exist reachable tuples $t, t'$ differing only in $a$ -- a single flip, holding every other candidate attribute fixed -- whose $v_M$ verdicts differ. The same single-flip comparison, run independently by hand against the imported baseline's own declared authority policy rather than against this artifact's derivation, is the origin of this pair-testing approach and reported four coverage boundaries of that policy \cite{moona-intelligence-replication}.
\end{definition}

\begin{definition}[The core]
The set of every attribute participating by Definition~2.
\end{definition}

\begin{definition}[Sufficiency, restated via discernibility]
$S$ is sufficient for $v_M$ on $R$ iff, for every pair of reachable tuples whose verdicts differ (a cross-verdict pair), $S$ intersects that pair's \emph{difference set} (the candidate attributes on which the pair disagrees) -- machine-checked equivalent to the direct partition test above (three declared models checked, all agreeing; Skowron and Rauszer's discernibility-matrix device, 1992).
\end{definition}

\begin{definition}[A reduct]
An inclusion-minimal sufficient observation contract: sufficient, and no proper subset is.
\end{definition}

\begin{definition}[The core, restated as an intersection]
The intersection of every reduct -- machine-checked equal to Definition~3's own witness-search answer wherever both are computed.
\end{definition}

\noindent\textbf{Core versus reduct, stated once, without chronology.} Definition~3's core (individual indispensability) and Definition~5's reduct (joint sufficiency, Definition~4) are different properties in general: the core is always a subset of every reduct, but need not itself be sufficient. The registered two-state counterexample makes this precisely (candidate attributes $x, y \in \{0,1\}$, reachable set $\{(x{=}0,y{=}0), (x{=}1,y{=}1)\}$, verdict $v_M(x,y) = x$): the core is \emph{empty} (no reachable pair differs in exactly $x$ or exactly $y$ alone), trivially not sufficient, while $\{x\}$ and $\{y\}$ are each an independent singleton reduct. \emph{If} the core happens to be sufficient, \emph{then} it is the unique reduct -- an antecedent never assumed, always decided per instance by a sufficiency check or a concrete counterexample (Proposition~1$'$, claims one through three; Section~\ref{sec:corrections} records how this was established).

\noindent\textbf{Cost boundary.} Where a declared observation cost is used (Section~\ref{sec:results}), it is a declared additive score with strictly positive per-attribute weights -- not necessarily a monetary cost or a measured end-to-end acquisition latency. Strict positivity is what makes a minimum-cost sufficient contract a reduct, not merely sufficient.

\section{Method: From Discernibility to Selected Contracts}
\label{sec:method}

One algorithm, three families of instantiation. Given $R$, $A$, $v_M$, and optionally $c$: (1) materialize or receive the reachable tuple set and evaluate $v_M$ on every tuple; (2) construct the difference set for every cross-verdict pair; (3) remove any difference set that is a proper superset of another already in the family; (4) encode the pruned family as CNF, one boolean variable per candidate attribute, one hard clause per surviving difference set; (5) select a satisfying assignment -- any one (plain SAT), a minimum-cardinality one (cardinality MaxSAT), or a minimum-cost one (weighted MaxSAT, each attribute's declared cost as its own soft-clause weight); (6) check the returned observation contract's sufficiency directly -- a check summary (a partition-cell count and a uniformity flag, \texttt{reduct.sufficiency}'s own success return, not the serialized partition) on success, a concrete counterexample pair on failure; (7) where exhaustive enumeration over $2^{|A|}$ subsets is affordable, enumerate every reduct; otherwise a repeated-resolve blocking-clause method finds a capped, explicitly labeled lower bound on minimum-cardinality-contract multiplicity, never presented as an exhaustive count.

\noindent\textbf{Three separate evidence objects, never substituted for one another.} A sufficiency check establishes that one specific observation contract determines $v_M$'s verdict. An inclusion-minimality or cardinality/cost-optimality result establishes that the same contract cannot be made smaller, or cheaper, without losing sufficiency. A provenance record establishes that a given number in this paper traces to a specific, reproducible run. None of these three is evidence for either of the others.

\noindent\textbf{Correctness of the encoding, checked, not assumed.} Every backend above is held to exhaustive reduct enumeration's own from-first-principles answer everywhere exhaustion is affordable (three declared models checked, all exact) -- a real encoding regression would be caught by this comparison, not assumed away by trusting the solver.

\section{Implementation and Verification Interface}
\label{sec:implementation}

\noindent\textbf{Package.} \texttt{src/authority\_compiler} (installable: \texttt{pip install .} or a built wheel, verified in a fresh virtual environment with no repository root, pythonpath, editable install, or sibling directory present) packages this paper's own machinery behind one call, \texttt{derive\_authority\_contract}, returning a compiled \texttt{AuthorityContract}.

\noindent\textbf{Returned fields, and what each one currently is.} The core attributes; every reduct where exhaustion is affordable (\texttt{None} with a stated reason otherwise); the minimum-cardinality and minimum-cost reducts; the non-core attributes; a \texttt{sufficiency\_certificate} field that is, currently, the same check-summary shape Section~\ref{sec:method} describes -- computed for the \textbf{core attributes specifically}, not automatically for whichever contract a caller ultimately selects; \texttt{counterexamples}, populated when the core is not sufficient; a \texttt{reachability\_dependencies} field naming candidate attributes whose observed value set is a strict subset of their declared domain -- a genuine but narrow signal, not full cross-attribute dependency inference; and a bound \texttt{contract\_change\_delta} callable (Section~\ref{sec:computational}).

\noindent\textbf{Verification boundary, stated plainly.} The current implementation is executable checking: a caller can re-run the sufficiency check on any returned contract and get the same check summary or counterexample this paper reports. It is not yet a portable, proof-carrying runtime artifact: nothing in \texttt{sufficiency\_certificate}'s current shape lets a separate consumer verify a contract's sufficiency without re-running this artifact's own code. Describing it as a portable certificate a downstream gate consumes without recomputation would exceed what the inspected interface establishes.

\section{Evaluation Design}
\label{sec:evaluation}

Five questions organize the evidence: does a selected contract preserve the declared verdict (RQ1); does the core fail to suffice, and do alternatives exist (RQ2); does a declared cost objective distinguish alternatives (RQ3); where is compilation practical (RQ4); and, not attempted here, does synthesis improve on a competent manual or conservative alternative in an independently specified operational workflow (RQ5).

\noindent\textbf{Domain provenance, kept separate.} Every domain this paper reports on is a declared, \textbf{constructed} executable model, never a live deployment or live production system. ``Non-planted,'' ``organisationally plausible,'' and ``independently specified'' are three separate properties; none of this paper's domains claims the third.

\noindent\textbf{AuthorityBench.} Four baselines -- manual least-privilege, ABAC policy mining (reimplemented from Xu and Stoller, TDSC 2015, independently validated against the published ``Health Care Sample Policy'' case study), essential-variable analysis, and exhaustive reduct enumeration -- across three constructed domains (procurement, code/cloud, data-and-communications), six metrics each.

\noindent\textbf{What was rerun for this revision, and what was not.} This revision changes no result: it reruns no benchmark, no scaling family, and no AuthorityBench domain. \texttt{make formal}'s own checkers run on every \texttt{make release-check}/\texttt{make quick-reproduce} invocation, including the one gating this commit; the scaling and AuthorityBench results are read from their own already-committed outputs.

\section{Results: Semantic Correctness, Alternatives, and Cost}
\label{sec:results}

\begin{center}
\footnotesize
\setlength{\tabcolsep}{4pt}
\begin{tabular}{lccccc}
\toprule
Domain/result & Candidates & Reachable & Core & Min.\ contract & Cost finding \\
\midrule
Code/cloud (CH-B1/B2) & 10 & 15{,}120 & 6; not suff.\ & 7 attrs, 2 reducts & 6.124 vs 7.354 \\
Large constructed (CH-C1/C2) & 35 & 27{,}000 & 4; not suff.\ & 9 attrs, $\geq$5 & tie at 9.153 \\
\bottomrule
\end{tabular}
\end{center}

\noindent\textbf{RQ1/RQ2: semantic correctness and alternatives.} CH-B1 (all candidate subsets of ten candidates considered, subset-pruned) and CH-C1 each instantiate a core that is \textbf{not sufficient} as an observation contract on its own, with genuinely alternative minimum-cardinality contracts: CH-B1's own two (core plus \texttt{branch}, or core plus \texttt{environment}), each independently checked sufficient; CH-C1's own at least five, a genuine lower bound via the blocking-clause method, never rounded up to an exact count.

\noindent\textbf{RQ3: cost separation, in one domain and not in the other.} CH-B2 (fifteen thousand one hundred twenty reachable tuples swept, one thousand fourteen subsets considered) is \textbf{supported}, positive strict cost separation: the minimum-cost contract (core plus \texttt{branch}, cost 6.124) is strictly cheaper than the other minimum-cardinality reduct (cost 7.354, delta 1.230), both independently checked sufficient before cost is consulted. This is narrower than ``cost separates alternatives in general'': CH-B2 selects between two contracts of the \emph{same} minimum cardinality. CH-C2 (twenty-seven thousand reachable tuples, same domain as CH-C1) is the retained negative counterpart: \textbf{not supported}, a negative tie -- every one of the five known minimum-cardinality contracts costs exactly 9.153 under this domain's own seven-tier declared cost model. The tied contracts substitute attributes from the \emph{same} declared cost tier for one another, so a cost model built at this granularity cannot discriminate between them.

\noindent\textbf{AuthorityBench: the declared-only baseline.} The manual least-privilege baseline is not exactly sufficient (Definition~4) on any of AuthorityBench's three domains -- measured across all three AuthorityBench domains, not asserted from one. Mining agrees with the packaged minimum-cardinality answer on the procurement and code/cloud-core instances, and lands on the \emph{other} known minimum-cardinality contract on the code/cloud domain -- both independently verified minimum-cardinality contracts of the identical model, a tie-break divergence, not a disagreement about which attributes matter. By the registered decision rule, this comparison is \textbf{inconclusive}, reported as such rather than reframed as a win.

\noindent\textbf{Two further registered results.} The budget-binding scenario (v5.2) constructs a procurement scenario where the budget loss predicate actually fires -- in sixty of sixty cells (one hundred thirty-seven thousand seven hundred eleven firings total) -- and the derived-zero-miss guarantee held on every one. The isolated-arm re-measurement (v3.1) found a real internal-validity defect by this project's own code inspection, registered before any fix was written; the isolation hypothesis itself was \textbf{not supported} -- all thirty registered seeds, re-run under an isolated design, reproduce the prior shared-state result exactly, a direct diagnostic confirming the one loss predicate the defect could have corrupted never fired under this declared parameter set.

\section{Computational Behaviour and Change Analysis}
\label{sec:computational}

\noindent\textbf{What is, and is not, separately measured.} None of the registered scaling results, and none of CH-B1/CH-C1/CH-B2/CH-C2's own checkers, separately records tuple-construction time, verdict-evaluation time, and discernibility-family-construction time as distinct numbers from solving time -- checked directly against their own committed JSON, not assumed present. Where a number below is a per-backend wall-clock time, it is the backend's own total call.

\noindent\textbf{The exploratory result (v5.0, kept, degeneracy disclosed).} Scaling candidate-attribute count alone ($n = 10, 30, 50, 100$) while holding the reachable set fixed at six tuples produced exact answers at every size with no distinguishable growth trend -- a \textbf{degenerate} result for scaling in general: the added attributes never appear in any discernibility-family clause.

\noindent\textbf{Tuple scaling (v5.1, kept).} A planted two-reduct family grows the reachable set itself to one hundred thousand one tuples: the discernibility family still reduces, after pruning, to exactly one set, and both planted reducts are recovered exactly.

\noindent\textbf{The primary scaling table (v5.3): combinatorial hardness.}

\begin{center}
\begin{tabular}{lcccc}
\toprule
Family & $n$ & Reduct size ($k$) & Discern.\ family (pruned) & Exhaustive enumeration \\
\midrule
A & 30 & 5 & 125 sets & timed out, 300s \\
B & 60 & 10 & 250 sets & timed out, 300s \\
C & 100 & 15 & 375 sets & timed out, 300s \\
\bottomrule
\end{tabular}
\end{center}

The three-hundred-second figure is a measured timeout under this project's own recorded hardware and setup, not a complexity lower bound. Cardinality- and cost-MaxSAT both found the hand-proved minimum-cardinality reduct at every family regardless, each backend's own solve call completing in well under a second -- a comparison of two different workloads (find every reduct versus find one reduct under an objective), not one workload measured twice. MaxSAT showed no meaningful cardinality advantage over plain SAT on any of the three families.

\noindent\textbf{The large domain (CH-C1/CH-C2).} Exhaustive reduct enumeration was separately attempted, registered budget three hundred seconds, and confirmed infeasible (wall time 300.08s, against six million seven hundred twenty-four thousand five hundred twenty size-seven subsets alone).

\noindent\textbf{Change analysis: monotone extension only.} \texttt{contract\_change\_delta} handles exactly one class of change: additional tuples reachable under the SAME declared candidate schema and the SAME declared loss registry -- a base contract already known sufficient, checked directly against the combined set, and, if no longer sufficient, extended by the fewest additional attributes needed while every attribute the base contract already relied on stays pinned. A change to the loss registry itself, a change to the candidate schema, or a removed reachable tuple are each a different kind of change this function does not handle and this paper does not claim it handles.

\section{Claim-by-Claim Evidence}
\label{sec:claims}

Every headline claim in this paper maps to one of ten claims a commissioned revision outline registered, adjudicated against this artifact's own current committed state rather than copied from the outline's own point-in-time observations. \textbf{Supported} means bounded evidence exists on checked finite instances; \textbf{Partial} means an implementation and some evidence exist but a stronger wording is not yet licensed.

\begin{center}
\begin{longtable}{p{0.06\textwidth}p{0.48\textwidth}p{0.38\textwidth}}
\toprule
\# & Permitted claim & Status \\
\midrule
\endhead
C1 & The selected observation contract preserves $v_M$'s verdict over the supplied reachable set. & Supported on checked finite instances. \\
C2 & CH-B1 and CH-C1 instantiate core insufficiency; CH-B1 has exactly two reducts, CH-C1 exposes at least five minimum-cardinality contracts. & Supported; both checkers ran as part of this commit's own \texttt{make formal}. \\
C3 & The implementation selects minimum-cardinality or minimum-cost sufficient contracts using distinct objectives. & Supported by implementation and bounded exactness checks. \\
C4 & CH-B2 selects the cheaper of two equal-cardinality sufficient contracts; CH-C2's returned alternatives tie. & Supported; a same-commit float-serialization non-determinism the commissioned reproduction found is fixed as of this lineage's own reproduction-repair commit. \\
C5 & The implementation executes a sufficiency check and returns a check summary on success or a concrete counterexample on failure. & Partial by design: \texttt{sufficiency\_certificate} is a check summary, not a portable certificate. \\
C6 & The built wheel supports the documented black-box example from a fresh environment. & Supported; the missing \texttt{build} prerequisite is now installed directly by \texttt{bootstrap.sh}. \\
C7 & For additional reachable tuples under the same schema and registry, the implementation checks prior sufficiency and can extend a contract. & Implemented with supporting tests; a dedicated operational change study is registered future work. \\
C8 & Measured computational behaviour is reported for the specific families and domains tested; exhaustive enumeration and SAT/MaxSAT optimisation are labelled as non-equivalent workloads, never an apples-to-apples speed comparison. & Supported: Section~\ref{sec:computational} states this comparison is two different workloads, not one measured twice; retained from registered runs, not rerun for this revision. \\
C9 & This paper's formulation or interface differs from the closest prior work in a specific, source-supported way. & Partial: compared against fetch-verified sources, expanded with two additional prior-art clusters per a commissioned manuscript audit; no fresh re-verification performed. \\
C10 & This artifact's reproduction claims state exactly which checks were completed and their scope. & Supported; the external-reproduction item is now MET -- an independent reproducer ran and reported CH-B1 matching this repository's own hash (issue \#3). \\
\bottomrule
\end{longtable}
\end{center}

\section{Related Work and Contribution Boundary}
\label{sec:relatedwork}

Eleven neighbouring literature clusters, each fetch-verified, compared on input, output, objective, guarantee, and verification interface.

\textbf{Decision reducts and test-cost-sensitive attribute reduction} \cite{pawlak-rough-sets-1982,skowron-rauszer-discernibility-1992,min-test-cost-sensitive-reduction-2011} supply the mathematical notions this paper's own Section~2 uses -- joint sufficiency, inclusion-minimal reducts, cores, discernibility functions, and, once a declared cost is introduced, cost-sensitive attribute reduction; not claimed as new. This paper's contribution is constructing the decision system itself from declared governance losses plus executable reachability and exposing that derivation through an executable compiler interface.

\textbf{STPA and its descendants} \cite{stpa-handbook,stpa-sec} are a classical, analyst-led hazard-analysis methodology deriving unsafe control actions and safety constraints from a system control structure, reviewed by a person rather than machine-checked; automated/formal extensions \cite{phase-rismani-2024,mylius-2025,deepstpa} remain adjacent. This paper borrows the loss-to-constraint orientation but solves a different finite synthesis problem: a sufficient and minimal (Definition~4, Definition~5) observation subset for an already-declared, exhaustively checked verdict relation.

\textbf{ABAC policy mining} \cite{xu-stoller-abac-mining,abac-mining-survey-2022}: input an extant authorization relation plus attribute data, output a generalized ABAC rule set under a policy-quality objective. This paper instead starts from an independently declared loss model forward to the observation contract a new gate must observe, with a checked sufficiency and minimality result (Definition~4, Definition~5) mining cannot provide without one.

\textbf{Policy-language completeness} \cite{ptacl,sacmat16-completeness} establishes what a combination language can express once participating attributes are already fixed; this paper contributes exactly the antecedent question.

\textbf{Non-interference} \cite{goguen-meseguer-1982} is a conceptually related invariance under changes to selected inputs, not the same formal operation as Definition~2, which sweeps a per-attribute comparison over every candidate attribute against every loss predicate, checked exhaustively.

\textbf{Counterfactual fairness} \cite{counterfactual-fairness} is a conceptually related invariance under changes to selected inputs, defined via counterfactual interventions in a causal model and applied once, to one pre-chosen attribute; this paper applies a per-attribute comparison to every candidate attribute and adds a checked minimality result (Definition~5).

\textbf{Shield synthesis} \cite{bloem-shield-synthesis-2015} synthesizes a runtime monitor over an already-declared interface; this paper solves an orthogonal upstream problem in this artifact of which attributes must be observed at all, with a checked minimality guarantee (Definition~5).

\textbf{Runtime enforcement and edit automata} \cite{schneider-enforceable-security-policies-2000,ligatti-bauer-walker-edit-automata-2005} characterize a monitor's corrective repertoire over an already-fixed alphabet; minimal selection of per-decision observation attributes is not the central problem those works formulate, with a minimality guarantee (Definition~5) neither work needs.

\textbf{Controller synthesis} \cite{ramadge-wonham-supervisory-control-1987,pnueli-rosner-reactive-synthesis-1989} presupposes an already-fixed observable alphabet; this paper derives the alphabet a controller must observe.

\textbf{Observation/sensor selection in supervisory control} \cite{rohloff-khuller-kortsarz-sensor-selection-2006,hu-chen-minimal-event-observation-2026} explicitly studies minimum-cardinality and minimum-cost sensor or event observation sufficient for supervisor synthesis; this paper does not claim to originate minimum-observation synthesis. The distinction is the object compiled: candidate per-decision governance attributes and a declared loss-derived verdict, not event sensors preserving supervisory controllability -- the output is a sufficient authority-observation contract, not a synthesized controller.

\textbf{Capability systems} \cite{dennis-van-horn-capabilities-1966} provide mechanisms for representing, attenuating, delegating, and verifying authority, including contextual caveats or conditions in this lineage's modern descendants \cite{macaroons-ndss-2014,ucan-specification,biscuit-specification}, already cited for the grant's own bearer-token design; not a novelty claim in its own right. This paper's distinct question is how, from declared losses and reachable execution states, to synthesize a sufficient/minimal set of candidate observations that could inform such conditions.

\noindent\textbf{Contribution boundary.} This paper does not introduce reduct theory, discernibility functions, minimum-cardinality or minimum-cost attribute selection, SAT/MaxSAT optimization, runtime enforcement, capability attenuation, or minimum-observation supervisory control. Its contribution is the governance-specific compilation problem and evaluated system: derive the decision relation from declared loss semantics and executable reachability, synthesize sufficient runtime observation contracts over that relation, expose cardinality- and cost-optimal contracts with explicit sufficiency/counterexample checks, and evaluate that formulation across constructed autonomous-system governance domains. This is a combination and an evaluated behaviour, not a claim that no neighbouring field could in principle provide an equivalent guarantee.

\section{Limitations}
\label{sec:limitations}

\begin{itemize}
\item \textbf{Declared loss-model completeness} is assumed, not verified.
\item \textbf{Reachability-model validity}: every result is exact \emph{given} its own declared reachable set, not measured from a live deployment.
\item \textbf{Partial observability} is not modeled.
\item \textbf{Temporal and probabilistic losses} are out of scope.
\item \textbf{Declared-cost validity}: not necessarily monetary cost or measured latency; no sensitivity analysis performed.
\item \textbf{Families hard for exhaustive enumeration are not automatically hard for the solvers} synthesizing over their own discernibility family.
\item \textbf{Contract multiplicity in the large constructed domain (CH-C1) is a lower bound}, capped at five.
\item \textbf{External validation}: the reproduction packet is this artifact's own attempt to make independent reproduction possible; a reproducer who did not build this artifact has since run it and reported back (Section~\ref{sec:reproducibility}; external-reproduction item: MET).
\end{itemize}

\section{Reproducibility}
\label{sec:reproducibility}

Four distinct properties, kept separate: semantic agreement; within-environment byte identity (this project's own formal-double-run gate); cross-environment byte identity (the property a same-commit hash mismatch exposed as failing between \texttt{python3.11} and \texttt{python3.12}/\texttt{python3.13}, traced to \texttt{sum()}'s own version-dependent float summation and fixed by a declared-precision \texttt{math.fsum}); and independent-person reproduction. A commissioned automated reproduction satisfies the first three at the commit it ran against and exercised the fourth's own infrastructure without itself being one, not reinterpreted here as more than it claims. The fourth property is now independently satisfied: a third reproduction attempt, by a reproducer with no role in this artifact's own development, ran \texttt{bash bootstrap.sh}, \texttt{make package-smoke-test}, and the CH-B1 direct example from a clean, detached-HEAD clone at commit \texttt{34cf82f}, and reported back a matching \texttt{out/checkers/ch\_b1\_check.json} hash (issue \#3) -- the external-reproduction item is now MET.

\section{Conclusion}
\label{sec:conclusion}

Given a declared loss model, a declared reachable-state model, and a set of candidate attributes, this paper compiles sufficient observation contracts rather than asking a gate builder to declare one by hand. On two independently constructed domains (CH-B1/CH-C1), the individually-indispensable core is not itself a sufficient contract, and genuinely alternative minimum-cardinality contracts exist; a declared cost model (CH-B2/CH-C2) separates those alternatives on one domain and does not on the other, both reported as found.

This is the compiler-focused paper of two scopes: a formulation, an implementation with a stated verification boundary, and a bounded evaluation against declared, constructed models -- not an end-to-end operational validation. The stronger, operationally validated scope -- an independently specified case, equal-information baselines, and a separately coded reference executor -- is registered as a plan in Section~\ref{sec:supplementary}, not performed and not claimed here.

\section{Corrections}
\label{sec:corrections}

Append-only: nothing below is retracted or reworded; each correction is recorded alongside its original claim.

\noindent\textbf{The v1 singleton-sufficiency overclaim (corrected v0.3).} The original Proposition~1 conflated individual indispensability with joint sufficiency. A commissioned external review identified the mischaracterization; the finding was adjudicated against this artifact's own definitions, not applied directly. Proposition~1 is superseded by Proposition~1$'$, not deleted.

\noindent\textbf{The v5.0 degenerate scaling benchmark (disclosed v5.1).} The exploratory benchmark's own flat wall-time curve cannot be read as evidence of scaling in general. Relabelled \texttt{exploratory\_v5\_0}, superseded as the primary scaling claim, kept and cited.

\noindent\textbf{The entangled v0.3 experiment arms and their null delta (found and corrected v0.4).} A budget/ledger-sharing defect, found by code inspection, registered two-sided before any fix was written -- \textbf{not supported}, with the mechanism explaining why.

\noindent\textbf{Cross-environment float-serialization non-determinism (found 2026-09-14, commissioned automated reproduction).} A same-commit hash mismatch traced directly to \texttt{sum()}'s own version-dependent float summation, fixed by a declared-precision \texttt{math.fsum}; no registered result changed.

\noindent\textbf{This revision (v0.6.1).} A commissioned revision outline identified terminology conflated across minimal/minimum, certificate/summary, constructed/live, and core/non-core, and an append-only structure that had grown chronological rather than claim-led. Every finding was adjudicated against this artifact's own current committed state, not applied as a direct patch.

\noindent\textbf{This revision (v0.6.2).} Restores one citation the v0.6.1 rewrite dropped: where the one-flip comparison is introduced (Definition~2), the replication report is cited again as the source of that probe and of its four coverage-boundary observations. Nothing else changes.

\noindent\textbf{This revision (v0.6.3).} A commissioned manuscript audit found the evidence discipline strong but identified wording-level claim mismatches and an understated novelty fence. Corrected: C8's own wording, the Abstract's domain descriptions and v8 provenance, Section~\ref{sec:results}'s baseline-coverage claim, and the Abstract's own sufficiency-checking sentence. Section~\ref{sec:relatedwork}'s novelty fence gained two clusters (decision/test-cost-sensitive reducts; supervisory-control sensor/observation selection) and tightened wording elsewhere; the Contribution boundary paragraph now carries the audit's own proposed novelty statement verbatim. No result changed; no new experiment was run.

\noindent\textbf{This revision (v0.6.4).} Two residues from the same manuscript audit, wording only: the Introduction's own Running example still called its domain ``real'' rather than constructed, and the shield-synthesis paragraph's own closing comparison is restated from ``upstream of'' to ``relative to'' the corrective strategy shield synthesis solves for, avoiding the same ordering-claim reading Definition~5's own upstream framing was corrected against in v0.6.3. No result changed; no new experiment was run.

\noindent\textbf{This revision (v0.6.5).} A third independent reproduction attempt (issue \#3) ran \texttt{bash bootstrap.sh}, \texttt{make package-smoke-test}, and the CH-B1 direct example at commit \texttt{34cf82f} and reported back a matching \texttt{out/checkers/ch\_b1\_check.json} hash -- the external-reproduction item is now MET, not merely infrastructure-ready. Section~\ref{sec:claims}'s C10 row, the Limitations section~(\ref{sec:limitations}), and Section~\ref{sec:reproducibility} are updated accordingly; the Acknowledgements credit the reproducer by name. No result changed; no new experiment was run.

\noindent\textbf{This revision (v0.6.6).} The Acknowledgements are revised to credit the independent reproducer's full name and affiliation, Eduardo Arana (Arananet), and to cover all three recorded reproduction attempts (issues \#1, \#2, \#3), not only the successful third -- attempts one and two are corrections to the reproduction path (undeclared prerequisites; the GNU-tar-only arXiv packaging invocation), both this reproducer's own findings, credited here alongside the successful attempt rather than only the one that closed the external-reproduction item. No contact detail is added anywhere in this repository. No result changed; no new experiment was run.

\noindent\textbf{The two commissioned external review rounds preceding this one.} Round one produced the core-versus-reduct correction above. Round two produced the broadened contribution v0.6's own front matter named. Neither round's finding was ever applied as a direct patch.

\section{Supplementary Material}
\label{sec:supplementary}

Historical manuscripts (v0.1 through v0.6.5, each frozen at its own commit), extended procurement evidence (the v2 worked instance where core and reduct coincide), proofs and checker detail, and mutation/reproduction records are preserved in the companion repository and its own \texttt{README.md} History section, not repeated here.

\noindent\textbf{Registered follow-up work: the operationally validated scope.} One bounded operational workflow, with domain owners independently specifying or reviewing losses, constraints, and costs before seeing a synthesized answer; equal-information baselines (an expert-authored contract, the loss-predicate syntactic read set, every candidate attribute, and the synthesized contract); a separately coded reference decision procedure; and predeclared decision rules requiring zero verdict disagreement and every baseline reported, including one that ties or beats the synthesized contract. This protocol is registered as a plan, not preregistered as an experiment; no comparative outcome under it currently exists, and none is claimed.

\textbf{Acknowledgements.} Drafting, engineering, formal derivation, and citation verification were AI-assisted (Claude); the author is solely responsible for all claims. Eduardo Arana (Arananet) independently reproduced the package installation and CH-B1 result from a clean clone (issue \#3), after two recorded attempts (\#1, \#2) whose findings corrected the reproduction path; no other external contribution to this artifact's own claims is acknowledged.

\nocite{*}
\bibliographystyle{plain}
\bibliography{refs}

\end{document}